\documentstyle[aas2pp4]{article}

\lefthead{Tresse \& Maddox} 
\righthead{H$\alpha$ Luminosity Function and Star Formation Rate at
$z\sim 0.2$}

\begin{document}

\title{The H$\alpha$ Luminosity Function and Star Formation Rate at $z
\sim 0.2$}

\author{Laurence Tresse\altaffilmark{1} and Steve J. Maddox}
\affil{Institute of Astronomy, Cambridge, CB3 0HA, UK}

\altaffiltext{1}{Visiting observer with the Canada-France-Hawaii
Telescope, operated by the NRC of Canada, the CNRS of France and the
University of Hawaii.}

\begin{abstract}

We have measured the H$\alpha$+[N~II] fluxes of the $I$-selected
Canada$-$France Redshift Survey (CFRS) galaxies lying at a redshift
$z$ below $0.3$, and hence derived the H$\alpha$ luminosity
function. The magnitude limits of the CFRS mean that only the galaxies
with $M_B\gtrsim -21$ mag were observed at these redshifts. We
obtained a total H$\alpha$ luminosity density of at least $10^{39.44
\pm 0.04}$ erg s$^{-1}$ Mpc$^{-3}$ at a mean $z=0.2$ for galaxies with
rest-fame $EW(H\alpha+[N~II])\gtrsim 10$ \AA.  This is twice the value
found in the local universe by \cite{gal95}.  Our H$\alpha$ star
formation rate, derived from Madau (1997) is higher than the UV
observations at same $z$, implying a UV dust extinction of $\sim 1$
mag.  We found a strong correlation between the H$\alpha$ luminosity
and the absolute magnitude in the $B$-band: $M(B_{AB}) = 46.7 - 1.6
\log L(H\alpha)$.  This work will serve as a basis of future studies
of H$\alpha$ luminosity distributions measured from optically-selected
spectroscopic surveys of the distant universe, and it will provide a
better understanding of the physical processes responsible for the
observed galaxy evolution.
 
\end{abstract}

\keywords{galaxies: luminosity function, mass function - galaxies: evolution}

\vspace*{3cm}
\begin{center}
Accepted for publication in {\it Astrophysical Journal} 
\end{center} 

\section{Introduction}

Deep spectroscopic surveys recently lead to a major breakthrough in
our understanding of global galaxy evolution.  The Canada$-$France
Redshift Survey (CFRS) (\cite{lil95c}) clearly demonstrated an
evolution in the galaxy population up to a redshift $z\sim1$, which
has been confirmed by the Autofib Survey (\cite{ell96}). Data
combining observations from the Keck Telescope and Hubble Space
Telescope (\cite{ste96}) reach the star-forming galaxy population up
to $z \sim 3-4$.  Thus, a picture of the star-formation history has
emerged from these observations (\cite{mad96}), suggesting that the
peak of star formation is in the range $z = 1.3-2.7$.

The observed evolution of the galaxy population is closely related to
the history of the star-forming galaxies showing spectral emission
lines.  Careful analysis of their spectra give crucial information
about the physical processes occurring in these galaxies
(\cite{tres96}, \cite{ham97}). In the optical wavelength range, the
H$\alpha$($\lambda 6563$) emission-line flux is a direct tracer of
recent star formation.  Massive, hot, short-lived OB stars emit
ultraviolet (UV) photons, that ionize the surrounding gas to form an
H~II region, where the recombinations produce spectral emission
lines. Of the Balmer lines, H$\alpha$ is the most directly
proportional to the ionizing UV stellar spectra at $\lambda < 912$
\AA\ (see \cite{ost89} for a review), because the weaker Balmer lines
are much more affected by the equivalent absorption lines produced in
stellar atmospheres.  The other commonly observed optical lines such
as [N~II]$\lambda\lambda6548,6583$, [S~II]$\lambda\lambda6717,6731$
and [O~II]$\lambda3727$, [O~III]$\lambda\lambda4959,5007$ depend
strongly on the metal fraction present in the gas. They have higher
ionizing potential than the Balmer lines, thus depend also on the
hardness of the ionizing stellar spectra.  Therefore they represent
only indirect tracers of recent star formation.

The major factor that affects measurements of the true H$\alpha$
emission fluxes is interstellar extinction.  If galaxies are observed
at high galactic latitudes, extinction due to our own Galaxy is
negligible ($\sim 0.05$ mag), hence most extinction is intrinsic to
the observed galaxy.  Star formation takes place in highly obscured
regions, so extinction corrections introduce a major uncertainty in
estimating the star formation rate.  However, since optical
wavelengths are less obscured than UV wavelengths, H$\alpha$ should be
a fairly good estimator of recent star formation.

One has to keep in mind that in studies where spectra are obtained for
the whole galaxy, the H$\alpha$ flux from OB stars is produced by H~II
regions which are distributed throughout both the bulge and disk.
Thus the observed line flux has been diluted by the whole stellar and
interstellar content of the galaxy. Though line flux measurements
depend on the content of an individual galaxy, they allow us to
compare populations of galaxies at different cosmic epochs, and so
quantify the global physical processes of evolution.  Also, if an
active nucleus is present in the central region of the galaxy, then
the H$\alpha$ flux is not correlated directly to forming stars, but is
a mixture of both.  Line ratio diagrams can usually separate H~II
galaxies from active galaxies.  However, if the AGN ionizing spectrum
is soft, then the stellar ionizing flux can dominate, and it is more
difficult to separate them.  At $z<0.3$, the number of AGN-like
galaxies is roughly $10\%$ of the number of emission-line galaxies
(\cite{tres96}, \cite{sar96}), and so the non-stellar component is not
dominant in the global H$\alpha$ luminosity measurements.
Nevertheless at larger redshifts, this situation may change if there
is a global evolution in the nuclear activity of galaxies.

Overall, measuring H$\alpha$ fluxes in representative
optically-selected galaxy samples at different cosmic epochs is a
major step towards the understanding of the evolutionary processes
occurring within the galaxy population.  The CFRS provides a
preliminary sample for studies of recent star formation. Its wavelength
range ($4500-8500$ \AA) allows the measurement of H$\alpha$ fluxes in
galaxies at $z \leq 0.3$.  A full description of the survey has been
published in \cite{lil95a} for the photometry, in \cite{lef95} for the
spectroscopy, in \cite{cra95} for its completeness, and in particular
the spectral analysis of the low redshift sample is presented in
\cite{tres96} (CFRS-XII).

Section 2 describes the flux measurements and the H$\alpha$ luminosity
function.  Section 3 discusses the star formation rate at $z\simeq
0.2$. Our conclusions are presented in Section 4. 

\section{The H$\alpha$ luminosity function}

The CFRS spectroscopic sample at $z \le 0.3$ ($138$ galaxies) is a fair
representation of field galaxies with absolute magnitudes $-22 <
M_{B_{AB}} < -14 $ mag~\footnote[1]{We assumed $H_0=50$ km s$^{-1}$
Mpc$^{-1}$ and $q_0=0.5$ throughout the paper.}, and with $ \langle z
\rangle = 0.2086$ (see Figure~\ref{fig1}. Also see Section 2 in
CFRS-XII for a detailed description).  Because of the $I$-band
($\lambda_c = 8320$ \AA) selection, these galaxies have been selected
by light from the old stellar population, rather than on their young
stars.  Consequently, our sample is less sensitive to galaxies
undergoing strong recent star formation than $B$- or
$H\alpha$-selected surveys at low-$z$. The wide $I$ band of width
$2000$ \AA\ does not favour galaxies with strong H$\alpha$ emission
lines.

\subsection{The emission-line measurements} 

The spectral resolution is $40$ \AA.  $117$ out of the $138$ spectra
to $z = 0.3$ exhibit the blended emission line
H$\alpha$+[N~II]$\lambda\lambda6548,65683$ (hereafter
H$\alpha$+[N~II]).  We have measured the integrated fluxes, equivalent
widths and the corresponding $1\ \sigma$ errors for 110 blends
(H$\alpha$+[N~II]) with the package {\small SPLOT} under {\small
IRAF/CL}.  For $7$ spectra, the blended line could not be properly
measured, since it was at the edge of the observed spectral window; in
our statistical analysis, we considered them as not observed.  The $1\
\sigma$ errors in line flux are typically $10\%$, the detection level
ranges from $2.5$ to $100\ \sigma$. Thus these errors are not a
significant source of uncertainty in our final estimate of the
H$\alpha$ luminosity function.  Rest-frame equivalent widths (REW) are
within the range $5-385$ \AA, and the mean is $48$ \AA.  Only $6\%$
out of the H$\alpha$ emitters have $REW(H\alpha + [N~II]) > 100$ \AA.
In Figure~\ref{fig2}, we show that except for $3$ H$\alpha$ emitters,
the sample has a detection limit of REW at about $10$ \AA, and REW
measurements are almost all above $3\ \sigma$.  At $z \gtrsim 0.2$,
the line [O~II] can be detected.  REW([O~II]) are in the range $4-94$
\AA, with a mean of $14$ \AA. The CFRS spectra at $z\leq 0.3$ are
limited in REW by the poor spectral resolution, rather than the S/N of
the continuum because they are the result of $\sim$7 hours exposure
time.  The $21$ spectra with no detection of (H$\alpha$+[N~II]) do not
always show H$\alpha$ in absorption, but all of them have the
characteristics of absorption-line spectra: strong $4000$ \AA\ break,
CaH\&K, H$\beta$ in absorption, G-band, MgI and a very red continuum
($(V-I_{AB})\ge1$).

\subsection{The H$\alpha$ flux measurements}

To obtain the flux in H$\alpha$, we corrected the flux $f$(H$\alpha$
+[N~II]) from the contribution of the doublet [N~II] where
$[N~II]\lambda6583/[N~II]\lambda6548 \sim 3$.  Using results from
a spectral analysis of the Stromlo-APM survey within the same absolute
magnitude range as our emission-line galaxies ($-21<M_B<-14$), we
determined values of $N_2=1.33[N~II]\lambda6583/H\alpha$ ranging
from $0.15$ to $0.55$ according to the strength of REW(H$\alpha$
+[N~II]) (\cite{tres98}). This method is in agreement with the trend
of the average parameters of [N~II]$\lambda$6583/H$\alpha$,
EW(H$\alpha$), and \vr \ colors of the UCM emission-lines galaxies
(\cite{gal96}, Tables~1 \&~2).  Figure~\ref{fig3}b shows the values
we have taken.
 
Where possible we estimated the interstellar extinction at H$\alpha$ using the
H$\alpha$/H$\beta$ Balmer decrement.  However, it is difficult to
correct for reddening for all spectra because the $40$ \AA\ spectral
resolution and stellar absorption mean that H$\beta$ is not always seen in
emission.  We were able to measure the H$\beta$ integrated fluxes for $55$
spectra.  For those spectra where we could make a $3\ \sigma$ measurement
for H$\beta$ ($40$ spectra), the extinction $C$ has been measured in
the following way.
$$10^{-C(-0.323)} = \frac{1}{2.86}\
\frac{f(H\alpha+[N~II])/(1+N_2)}{f(H\beta)} $$ The H$\alpha$/H$\beta$
intensity ratio ($2.86$) is for case B recombination with a density of
$100$ cm$^{-3}$, and a temperature of $10\ 000$ K (\cite{ost89}).
$[X(H\alpha)-X(H\beta)]/X(H\beta) = -0.323$ is the value given by the
extinction law $X(\lambda)$ taken from \cite{sea79}.  We point out
that although the different extinction laws are different in the UV,
they behave similarly in the visible, hence our results are
independent of the choice made.  For these $40$ spectra, $\langle C
\rangle = 0.36$.  For the other $15$ spectra with H$\beta$ in emission,
the average $C$ is $0.82$, or $0.50$ if a $2$ \AA\ stellar absorption is
accounted for at H$\beta$; we retained the latter value.  For $10$
spectra, we could not measure H$\beta$ because of a sky line near it,
and for the $44$ remaining, H$\beta$ is not detected in emission,
probably because of stellar absorption.  For these $54$ spectra, we
decided to apply a correction of $C=0.45$, corresponding to a
reddening parameter $A_V = C\ R/1.47 \simeq 1$ mag ($R=3.2$,
\cite{sea79}), which is the average value measured by \cite{ken92} in
nearby emission-line galaxies.  We then corrected our H$\alpha$ fluxes
for reddening as follows:
$$ \frac{f(H\alpha+[N~II])}{1+N_2} \ 10^{\ C (1-0.323)}\ .$$ 
We note that the final H$\alpha$ luminosity densities using these
reddening corrections are not significantly different to that using an
average of $A_V= 1$ mag for all galaxies.

The $1\farcs 75$ spectral slits of the CFRS did not always contain the
whole galaxy. To have consistent data, we corrected our fluxes by an
aperture correction. For this, we integrated the spectral flux in the
V band, and transformed it into magnitudes.  The spectrophotometric
flux calibration across the spectra is generally accurate to better
than $10\%$ (see e.g. Figure 9 in \cite{lef95}).  The $V$ magnitudes from
the CFRS imaging are given by $I_{AB} +(V-I)_{AB}$, where $I_{AB}$ is
the isophotal magnitude, and $(V-I)_{AB}$ is the color of the galaxy
measured in an aperture of $3\arcsec$.  The aperture correction is $a
= V_{image} - V_{spectrum}$. For our emission-line galaxies, the
values range between $0$ and $1.6$ mag, with an average of $0.52$.
The dereddened H$\alpha$ flux is then aperture corrected to give the
final estimate of the H$\alpha$ fluxes:
$$ f(H\alpha) = \frac{f(H\alpha+[N~II])}{1+N_2} 10^{\ C (1-0.323)}
10^{\ 0.4\ a}. $$ 
Since at $z\simeq 0.26$ H$\alpha$ falls at the center of our $I$ band
($8320$ \AA), we could check these aperture corrections, as follows.
The integrated flux in erg s$^{-1}$ cm$^{-2}$ in the blended line
(H$\alpha$ + [N~II]) can also be computed using the observed 
EW(H$\alpha$+[N~II]) in \AA\ and the $I$-band magnitude in erg
s$^{-1}$ cm$^{-2}$ \AA$^{-1}$ to estimate the continuum flux such as,
$$f(H\alpha+[N~II]) = EW(H\alpha + [N~II])\ 1.5\ 10^{-9}\ 10^{-0.4\
I_{AB}}.$$ The differences between the two methods are small, of order
$13\%$. They are due to the facts that the color $(V-I)_{AB}$ should
be the one measured in the spectral slit, and not in a $3\arcsec$
aperture; and that the true continuum level at H$\alpha$ may not be
exactly the mean value given by the $I$-band magnitude.  These
discrepancies are much smaller than our Poisson errors, so they do not
affect our results on the H$\alpha$ luminosity function.  Another
point is that according Kennicutt (1983), the H$\alpha$ nuclear
emission (H~II region complexes in the inner disk and bulge regions)
is, in general, rarely significant in comparison with the H$\alpha$
emission from the whole galaxy. Hence, our aperture corrections should
not overestimate the H$\alpha$ measurements, even in the cases 
of starburst nuclei.  

Finally, the H$\alpha$ luminosity in erg s$^{-1}$ is given by: 
$$ L(H\alpha) = 4\ \pi \ (3.086\ 10^{24}\ d_L)^2\ f( H\alpha), $$ 
where $f$(H$\alpha$) is the integrated flux in erg s$^{-1}$ cm$^{-2}$, $d_L$
is the luminosity distance in Mpc.

\subsection{Calculation of the H$\alpha$ luminosity function}

Figure~\ref{fig4} shows the comoving H$\alpha$ luminosity density
estimated from our raw data, and from the data after applying each of
the corrections described in Section 2.  These densities were obtained
using the $V_{max}$ formalism, i.e.,
$$\Phi [ \log{L(H\alpha)} ]\ \Delta \log{L(H\alpha)} = \sum_{i}
\frac{1}{V_{max}^i}.$$ 
$V_{max}^i$ is the comoving volume in which galaxy $i$ can be detected
in this $I$-selected survey ($17.5<I_{AB}<22.5$), and the sum is over
galaxies with H$\alpha$ luminosity within the interval
$[\log{L(H\alpha)} \pm 0.5\Delta \log{L(H\alpha) }]$.
$$ V_{max} = \int_{max(z=0,I_{AB}=17.5)}^{min(z=0.3,I_{AB}=22.5)}\
d_{<}^2\ \frac{dr}{dz}\ \Omega \ dz,$$ where $d_{<}^2$ is the angular
distance, $dr$ the comoving distance at $z$, and $\Omega$ is the
effective solid angle of the CFRS observations; $5$ fields of
$10\arcmin^2$ individually weighted by the number of spectroscopic
observations out of the photometric observations.  We plotted the
comoving densities at the barycenter of the $\log{L(H\alpha)}$ of the
$N_i$ data belonging to the interval $[\log{L(H\alpha)} \pm 0.5\Delta
\log{L(H\alpha) }]$.  The density error bars are Poisson errors; $\log{
(1 \pm 1/ \sqrt{N_i} )} $.  Figure~\ref{fig4} shows also the Schechter
(1976) fit to the local H$\alpha$ luminosity function (LF) measured by
\cite{gal95} ($\alpha=-1.3$, $\phi^{*}=10^{-3.2}$ Mpc$^{-3}$,
$L^{*}=10^{42.15}$ erg s$^{-1}$).

Note that we used the $V_{max}$ based on the $I_{AB}$ magnitudes,
since the galaxy selection is based on this measurement. For the
lowest REW, our measurements are not absolutely complete, and so we
should, in principle, include this effect in our estimates of
$V_{max}$. In practice, there is a strong correlation between
L(H$\alpha$) and $M_B$ (see Figure~\ref{fig5}b), which means that
$V_{max}$ should be only slightly smaller in these cases. We expect
the difference in $V_{max}$ to be small, since lines with low REW are
observable even at our maximum redshift, $z=0.3$, as can be seen in
Figure~\ref{fig3}d.  An extreme upper limit to this effect can be
obtained by artificially setting $EW(H\alpha+[N~II])=5$ \AA\
(roughly $+1 \sigma$), or $1$ \AA\ for our non H$\alpha$-emitting 
galaxies (15\% of the sample).  Then using the I$_{AB}$ magnitudes, we
can obtain an approximate flux in H($\alpha$)+[N~II], that we also
corrected for N$_2$, C (taken as 0.45), and aperture as described in
Section 2.2.  The LF we obtained in both cases is within the Poisson
error bars of our actual results.  As these $21$ galaxies are very
red, a small EW does not produce a small L(H$\alpha$), so they do not
produce a steeper LF at the faint end. Thus, our estimated LF is quite
stable.

We fitted our overall H$\alpha$ LF in Figure~\ref{fig4}d, with a
Schechter function, and the best-fitting parameters given by a
weighted minimum $\chi^2$ are $\alpha = -1.54 \pm 0.08 $, $\phi^{*} =
10^{-3.28 \pm 0.15}$ Mpc$^{-3}$, $L^{*} = 10^{42.50 \pm 0.23 }$ erg
s$^{-1}$.  If we exclude the faintest bin from the fit which contains
only four galaxies, we find: 
\begin{eqnarray*}
\alpha & = & -1.35 \pm 0.06  \\
\phi^{*} & = &10^{-2.83 \pm 0.09}\ Mpc^{-3} \\
L^{*} & = & 10^{42.13 \pm 0.13}\ erg\ s^{-1} \\ 
\end{eqnarray*}
The quoted errors are the formal $\chi^2$ fit standard deviations,
assuming the parameters are uncorrelated.  In fact, these three
parameters are highly correlated, so the formal errors are not
realistic estimates; the difference between these two fits give a 
realistic estimate of the true errors. Figure~\ref{fig1} shows that
at different redshifts, our survey does not sample exactly the same
range of absolute magnitudes.  In principle, this could affect our LF,
so we calculated the H$\alpha$ LF using only the galaxies at
$0.17<z<0.3$, and $-18<M(B_{AB})<-21$, i.e. where the data constitute
an homogeneous subsample. The result is shown in Figure~\ref{fig5}a
and we can see that the resulting densities are entirely consistent
with our fitted LFs.

Our H$\alpha$ LF at $z\leq 0.3$ samples fainter H$\alpha$ luminosities
than the \cite{gal95} LF; their lowest data is above $\log L(H\alpha)
= 40.4$ erg s$^{-1}$, ours is at $39.2$ erg s$^{-1}$.  On the other
hand, the ($I_{AB}<17.5$) CFRS limit and ($z<0.3$) cut lead to a small
volume where we sample bright galaxies with $M_{B_{AB}} < -21$ mag.
According to Figure~\ref{fig5}b, this corresponds to $\log L(H\alpha)
\gtrsim 42.3$ erg s$^{-1}$ which explains the large statistical error
on our brightest H$\alpha$ luminosity bin.  Also our sample contains a
larger proportion of emission-line galaxies with $5 < $REW(H$\alpha$ +
[N~II]) $< 30$ \AA. This can be seen in comparing Figures~9 \&~10 in
\cite{gal96}, and our Figures~\ref{fig3}c and \ref{fig3}d.

Although it may be expected that blue galaxies correspond to
star-forming galaxies, Figure~\ref{fig5}c shows that there is not a
clear correlation between the $(V-I)$ color of a galaxy and
$L$(H$\alpha$) in our sample.  There is however, a trend seen in
Figure~\ref{fig5}e, implying blue galaxies tend to have higher
REW(H$\alpha$ + [N~II]). The latter is in agreement with Figure~10 in
Kennicutt (1983), which plots REW(H$\alpha$ + [N~II]) against \bv \
colors for a nearby sample of Sa-Irr galaxies. These figures show that
REW(H$\alpha$) is sensitive to the ratio of ionizing stars (which
produce H$\alpha$) to lower mass red giant stars (which produce the
stellar continuum at H$\alpha$).

On one hand, this means that we have detected red~\footnote[1]{At $z
<0.3 $, galaxies having $(V-I)_{AB} \gtrsim 0.7$ are redder than a
local Sb spiral (see Figure~13 in \cite{tres96})}
galaxies which show both low and high
H$\alpha$ fluxes, and that these contribute to the overall
luminosity as well as the blue galaxies, for which emission lines are
easier to detect because of a lower stellar continuum at H$\alpha$.
Following the terminology in \cite{gal96}, these red galaxies are
likely to correspond to Starbust Nuclei Galaxies (SBN) if H$\alpha$ is
bright, and to Dwarf Amorphous Nuclear Starbursts (DANS) when $\log
L(H\alpha) < 41.6$. 
Both classes are spiral galaxies.  According to
\cite{tres96} and \cite{gal96}, they represent $\sim 40\%$ of
emission-line galaxies, the remaining being spectrally classified as
H~II galaxies, blue compact galaxies, or active galaxies.

On the other hand, the lack of a correlation between H$\alpha$
luminosities and continuum colors means that H$\alpha$ production is
independant of the overall stellar content, and depends only on the
recent star formation.  The observed $(V-I)_{AB}$ at $0.1<z<0.3$ is
rest-frame $(B-R)_{AB}$ which may be significantly reddened.  However
the typical reddening is $(A_B-A_R) \sim 0.2$ mag which would not make
blue galaxies appear as red as we observe.  The independence of
L(H$\alpha$) on total stellar content can also be seen in the fact
that red galaxies classified in class (B) by \cite{tres96} often show
spectral features proving the presence of a dominant old stellar
population (CaH\&K, G-band, MgI; see Figure~1 of \cite{tres96}).  So,
we suspect that some red galaxies show recent star formation similar
to blue galaxies.

Figure~\ref{fig5}b shows a tight relation between $H\alpha$
luminosities and $B$ luminosities, both of which being closely related
to recent star formation. 
A least squares fit gives:
$$ M(B_{AB})= 46.7 - 1.6 \log L(H\alpha). $$ If this relation holds at
higher redshifts, $0.3 <z<1$, we would expect that the high-$z$ CFRS
galaxies, which are brighter than $-20$ in $B$, should exhibit
stronger H$\alpha$ emission lines than those observed locally. This is
in agreement with the results of \cite{ham97} who found a large
increase in the comoving [O~II] luminosities up to $z\sim 1$ in the
CFRS ([O~II] luminosities being indirectly related to H$\alpha$
luminosities).

\section{Star Formation rate at $z\simeq 0.2$} 

We integrated our best-fit luminosity function to give the overall
comoving luminosity density at $\langle z \rangle = 0.2$:
$${\cal L}(H\alpha) = \int_{0}^{\infty} \phi(L) \ L\ dL = \phi^{*} \
L^{*} \ \Gamma(2+\alpha) $$ We then find a total H$\alpha$ luminosity
per unit volume ranging from $10^{39.44 \pm 0.04 }$ to $10^{39.50 \pm
0.07 }$ erg~s$^{-1}$~Mpc$^{-3}$ at $z\simeq 0.2$ from our two previous
LF best fits. The errors quoted here are the standard deviations
taking into account that the three Schechter parameters are
correlated, and so should be realistic estimates.  Our value is $2.2$
to $2.6$ times higher than the value of \cite{gal95} at $z\sim 0$
($10^{39.09\pm0.04}$, see Errata, 1996, \apj, 459, L43).
Consequently, our result shows that the star formation rate (SFR) is
higher at $z\simeq 0.2$ than that found in the local universe for
galaxies with $REW(H\alpha+[N~II])\gtrsim 10$ \AA.  Figure~\ref{fig6}
shows our data and the local measurements in the SFR vs. z plot from
Madau et al. (1997) assuming a Salpeter (1955) IMF including stars in
the mass range $0.1 < M < 125$ M$_{\odot}$. We used their conversion
${\cal L}(H\alpha) = 10^{41.15}\ SFR/M_{\odot}\ yr^{-1}$, based on
the stellar population synthesis models of \cite{bru97}.  The other
points in this plot are from rest-frame UV continuum measurements.
The comparison between the UV and H$\alpha$ results is not
straightforward for the following reasons.  As noted in the
introduction, H$\alpha$ fluxes come from ionized gas surrounding OB
stars, so are directly correlated to short-lived stars, once dust
corrected with the Balmer decrement.  The UV continuum comes from both
short- and long-lived stars; the long-lived (late B, A0) stars
contribute more to the UV continuum at larger UV wavelengths. In
addition, \cite{cal97} pointed that longer-lived non-ionizing stars
are likely to be found in less obscured regions, than ionizing
stars. This agrees with observations of starbursts where the
extinction obtained with the Balmer decrement is found to be higher
than the extinction obtained from UV continuum (\cite{kee93}).  Given
that UV photons are more obscured than H$\alpha$ photons, we expect
that UV data should be less correlated to short-lived stars than
H$\alpha$ data.  Overall, converting H$\alpha$ luminosities to SFR
depends less sensitively on the dust correction than on the assumed
IMF. Converting UV luminosities to SFR depends less sensitively on
IMFs in a standard cosmology (\cite{bau97}), than on uncertain dust
extinction.

In Figure~\ref{fig6}, the H$\alpha$ data are reddening corrected,
while UV data are not, thus H$\alpha$ data must give higher values of
star formation rate, than their counterpart in the UV.  \cite{cal94}
measured an {\it effective} dust extinction law from a sample of
extended regions such as the central regions of starburst and blue
compact galaxies.  Their law is characterized by the absence of a 2175
\AA\ dust feature.  Assuming $A_V=1$ mag, it predicts an average dust
extinction at $2000$ \AA, that is $1.95$ times more in flux than at
H$\alpha$.  \cite{trey98} have a preliminary measurement of the
UV($2000$ \AA) density at $z=0.15$, and find it is consistent with the
UV($2800$ \AA) CFRS data (\cite{lil96}).
According to the results from \cite{cal97}, our Balmer decrement
reddening of $A_V=1$ mag, would imply that the UV stellar continuum at
$2000$ and $2800$ \AA\ has an extinction of $1.3$ and $1$ mag
respectively. Thus, our SFR result at $\langle z \rangle = 0.2$ seems
to be consistent with the UV($2000$ \AA) data at $\langle z \rangle
= 0.15$, and the UV($2800$ \AA) data at $\langle z \rangle = 0.35$
within the error bars.

\section{Conclusion}

We constructed an optically-selected H$\alpha$ luminosity function at
$z \simeq 0.2$; this will be useful as a comparison to future
near-infrared spectroscopic surveys which will detect H$\alpha$ in
galaxies near the expected peak of SFR.  We find a total H$\alpha$
luminosity at least twice than the one measured in the local universe
by \cite{gal95} for galaxies with $REW(H\alpha+[N~II])\gtrsim 10$ \AA.
If the SFR evolution follows an $(1+z)^{3}$ law, then ${\cal
L}(H\alpha)$ should decrease by a factor $1.7$ from $z=0.2$ to $z=0$.
This factor is marginally outside the $1\ \sigma$ errors, and may
suggest that the local H$\alpha$ density is low by $\sim 20\%$. This
may correspond to the local under-density seen in optical redshift
surveys (see \cite{zucca97}).  If the local estimate is correct then
it implies an evolution proportional to $ (1+z)^{4.4}$.

Since the number of hydrogen ionizing photons ($\lambda < 912$ \AA)
emitted by a star is proportional to the H$\alpha$ recombination line,
the total flux in H$\alpha$ is a good tracer of the number of ionizing
stars within emission-line galaxies. The IMF introduces uncertainties
in the relation between H$\alpha$ luminosity and star formation rate;
H$\alpha$ traces only the massive, hot, short-lived stars, and
therefore assumptions on the remaining fraction of cooler stars have
to be made.  Also, since star formation takes place in highly obscured
regions, newly formed stars are not detected in the UV or optical
observations.  Taking the SFR factor conversions from Madau (1997),
our result is consistent with UV data corrected for $\sim 1$ mag of
dust extinction.  Larger dust corrections would imply either an IMF
with a shallower slope than the one from Salpeter, or an
underestimation of the total H$\alpha$ luminosity. However we are
aware that uncertainties in models and in UV dust extinction are still
large.

Another interesting result is the strong correlation observed between
the flux emitted in the rest-frame $B$-band and $H\alpha$
luminosity. Applying this relation to the CFRS where the rest-frame
$B$-band is directly observed at $z=0.9$, the range of sampled
magnitudes suggests that all emission-line galaxies should be strong
H$\alpha$ emitters.  This is in agreement with \cite{ham97} who find
an increase of the comoving number of [O~II] CFRS emitters by a factor
at least $5$ up to $z\sim 1$.  This correlates with the evolution seen
in the CFRS LFs, which is due to these bright, star-forming
galaxies. At high redshifts, deep spectroscopic surveys clearly become
biased toward strong emission-line, dust-free emitters, which
contribute to the increase of the total luminosity of these galaxies
at earlier epochs, however they may be only the tip of the iceberg of
all H$\alpha$ emitters.

\acknowledgments

It is a pleasure for LT to thank her colleagues David Crampton, Fran\c
cois Hammer, Olivier Le F\`evre and Simon Lilly, who made the CFRS
survey possible. This work has benefited from fruitful discussions
with Stephane Charlot.  We thank Piero Madau, who kindly provided his
most recent SFR conversion factors. LT acknowledges support from the
European HCM program/ERBCHBICT941612. SJM acknowledges support from a
PPARC advanced fellowship.

\onecolumn

\clearpage
\plotone{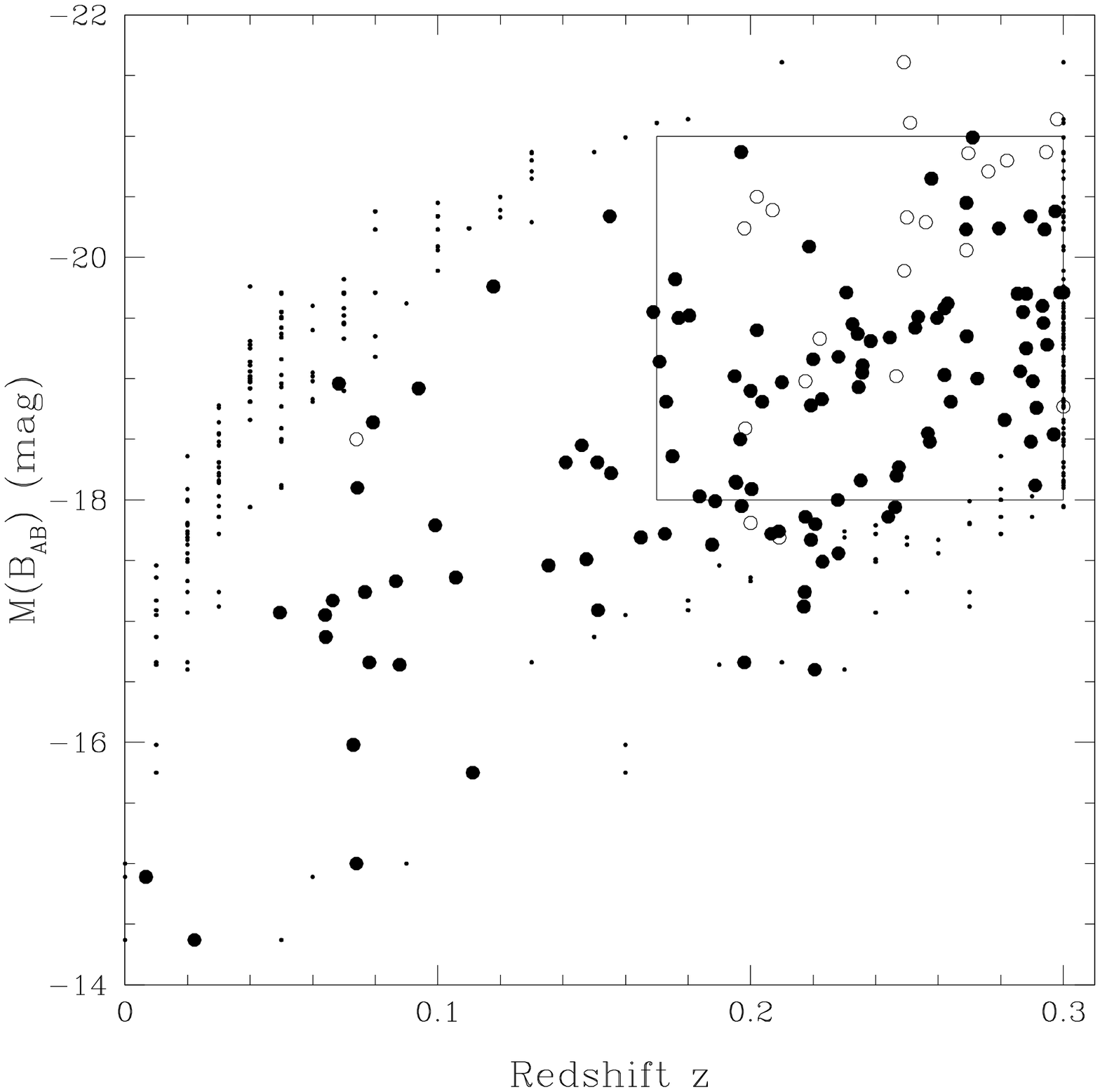}
\figcaption[fig1.ps]{Absolute magnitudes in $B_{AB}$ versus redshift $z$
of the 138 CFRS galaxies at $z\leq 0.3$. The filled circles represent
H$\alpha$ emitting galaxies, the open circles show H$\alpha$ absorbing
galaxies. The small dots are the minimal and maximal redshifts in
which a galaxy could be detected as used in the $V_{max}$ method (see
text). The rectangle shows an homogeneous subsample of galaxies at
$0.17<z<0.3$ and $-21<M(B_{AB})<-18$.
\label{fig1}}

\plotone{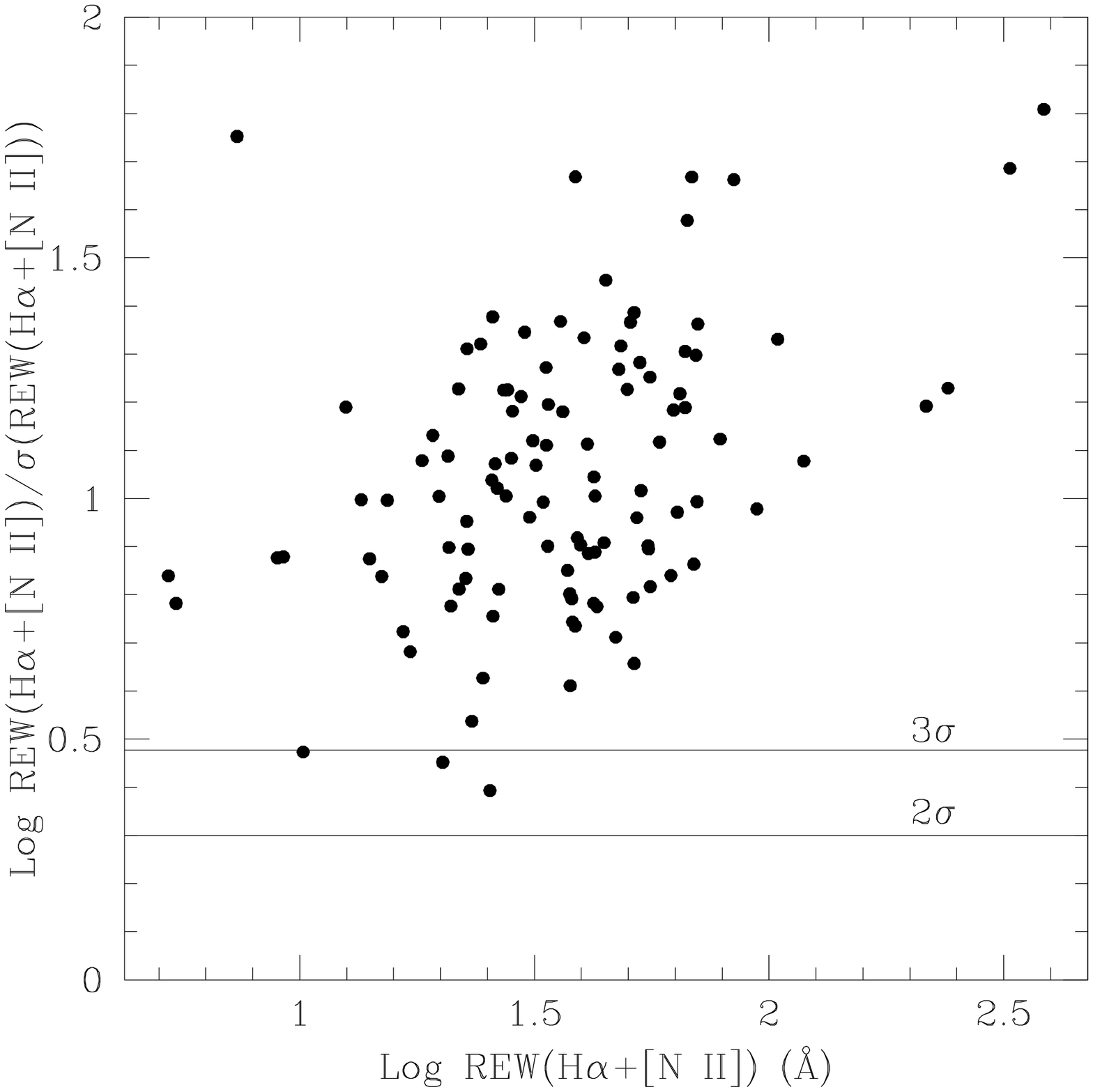}
\figcaption[fig2.ps]{ Detection level of REW(H$\alpha$+[N~II]) versus
REW(H$\alpha$+[N~II]). \label{fig2}}

\plotone{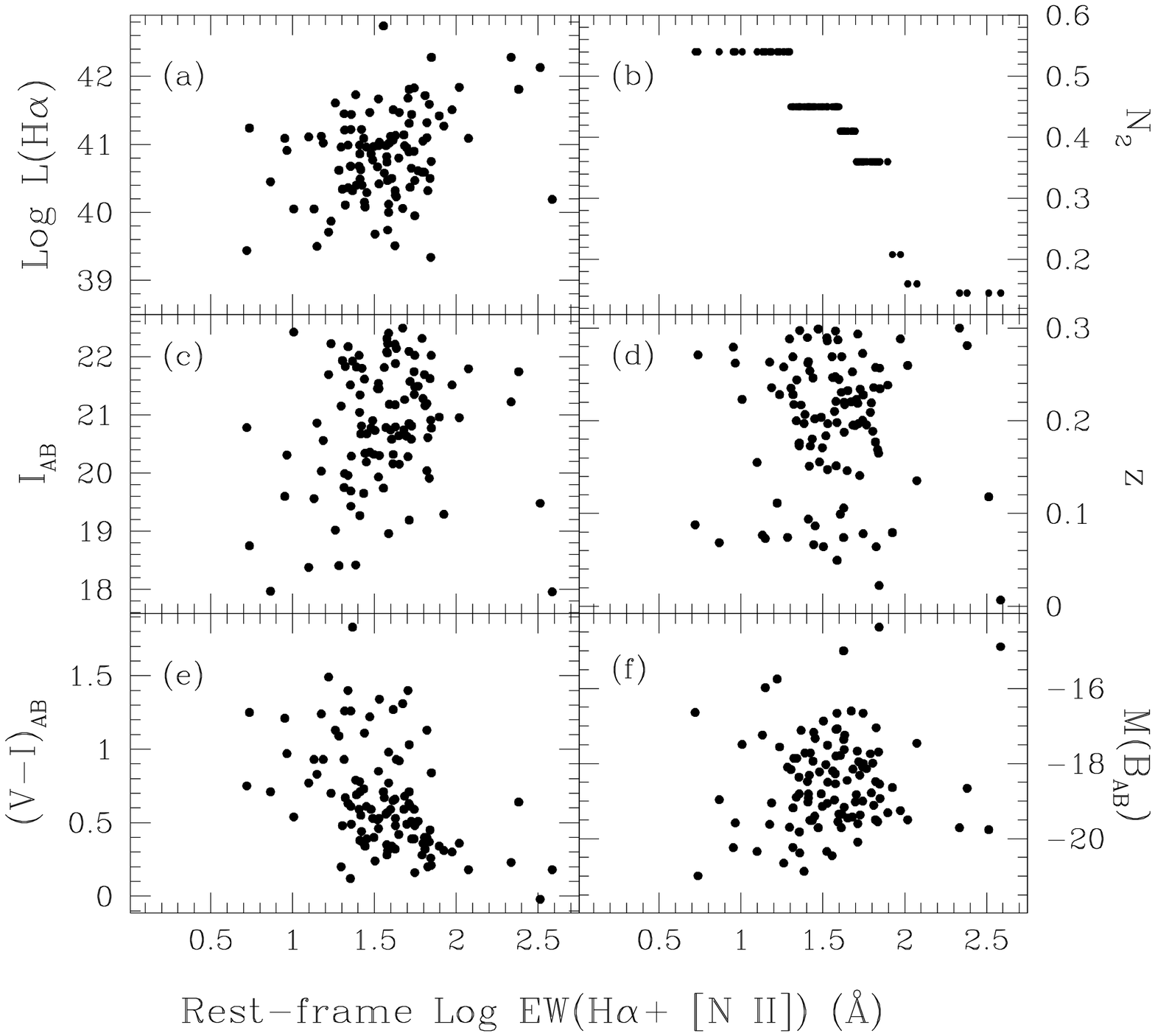}
\figcaption[fig3.ps]{Relations between REW(H$\alpha$+[N~II]) and
other parameters as labeled. They are useful to compare our sample
with other emission-line surveys. \label{fig3}}

\plotone{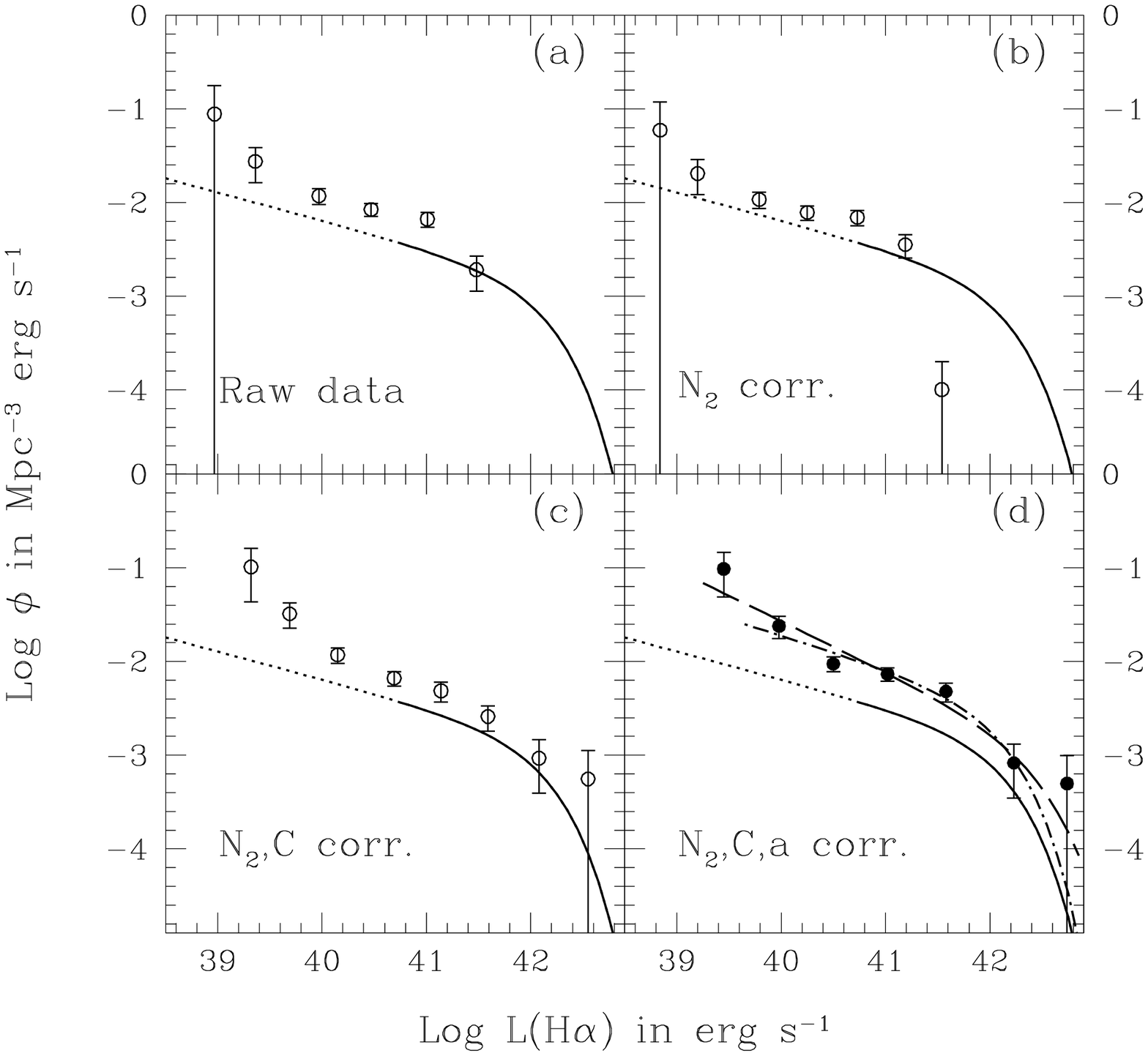}
\figcaption[fig4.ps]{Overall H$\alpha$ luminosity function at $z\le
0.3$. The LFs with open circles show the successive corrections to
obtain the final LF shown by the filled circles in panel (d).  The raw
data are shown in panel (a).  Panel (b) shows this after correcting
for the [N~II] contribution to H$\alpha$ ($N_2$). Panel (c) is further
corrected for the reddening $C$.  Applying the aperture correction $a$
gives the final LF in panel (d).  In each plot, the solid curve is the
Gallego et al. 1996 H$\alpha$ LF at $z\simeq 0$, the dotted curve is
its extension to fainter luminosities. The long-dashed and dot-dashed
curves are the best Schechter (1976) fits to our data, the latter one
is given by excluding the faintest bin. \label{fig4}}

\plotone{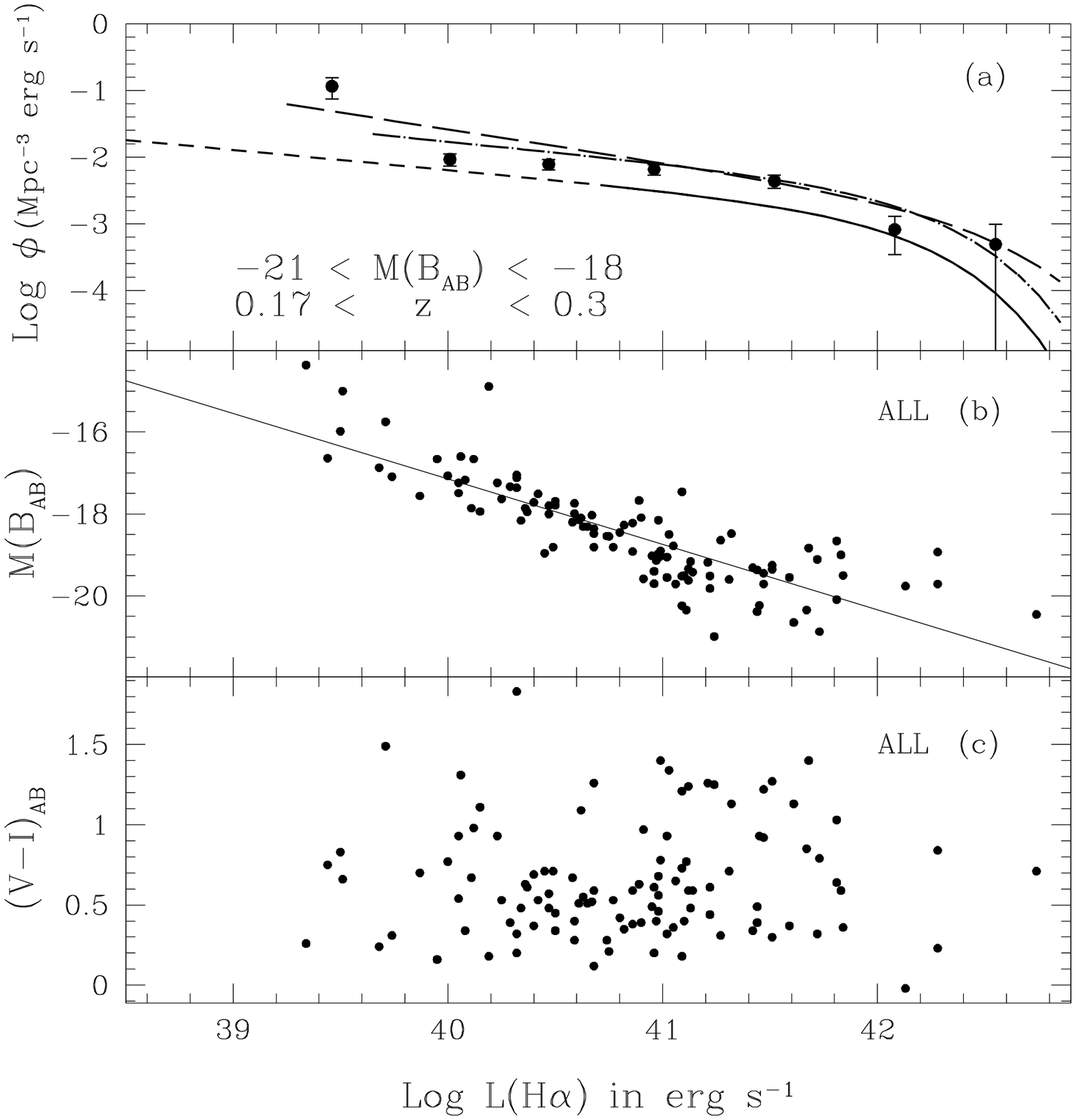}
\figcaption[fig5.ps]{The panel (a) shows the H$\alpha$ LF of the data
within the rectangle shown in Figure~\ref{fig1}. The curves are the
same as in Figure~\ref{fig4}d. The panels (b) and (c) show $M(B_{AB})$
and $(V-I)_{AB}$ versus H$\alpha$ luminosities of all the H$\alpha$
emitters.  Galaxies redder than a local Sb spiral have $(V-I)_{AB}
\gtrsim 0.7$ at $z<0.3$. \label{fig5}}

\plotone{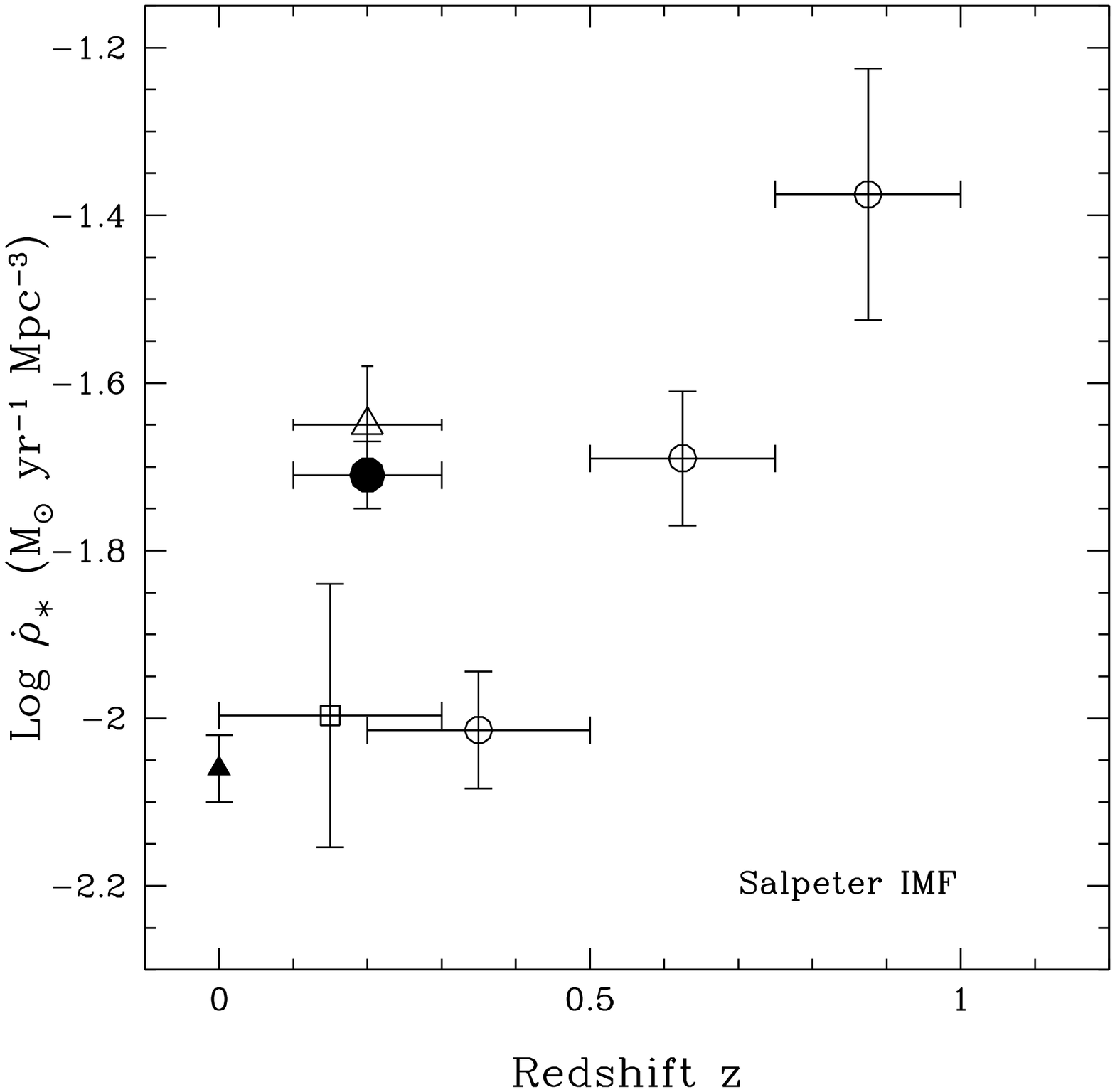}
\figcaption[fig6.ps]{Comoving volume-averaged star formation rate
versus redshift. The open circles are from the ``LF-estimated''
UV($2800$ \AA) CFRS data of \cite{lil96}, the open square is from
UV($2000$ \AA) data of \cite{trey98}, and the filled triangle is from
H$\alpha$ data of \cite{gal95}. The filled circle is our H$\alpha$
result (the open star is our result if the faintest bin of our
H$\alpha$ LF is included.) We used the Madau et al. (1997) conversion
factors; $\log {\cal L}(H\alpha) =41.15 + \log \dot{\rho_{\ast}}$, and $\log
{\cal L}_{UV} = 27.9 + \log \dot{\rho_{\ast}}$, assuming a Salpeter IMF and
including stars in the $0.1-125\ M_{\odot}$ mass range.
\label{fig6}}

\end{document}